\documentstyle[epsf,prl,aps,twocolumn]{revtex}

\begin{document}

\title{Comment on ``Experimental Demonstration of
Violations of the Second Law of Thermodynamics
for Small Systems and Short Time Scales''}

\author{S. Tasaki, I. Terasaki and T. Monnai}
\address{
Department of Applied Physics, 
Waseda University,
3-4-1 Okubo, Shinjuku-ku,
Tokyo 169-8555,
Japan}
\date{\today}
\maketitle

In a recent letter, Wang et al.\cite{WSMSE}
claimed that they experimentally demonstrate the 
violation of the second law of thermodynamics
for a colloidal particle in an optical
trap moving at a constant velocity. 
We show that their results cannot conclude the violation of 
the second law since
(i) the particle does not do the net work and (ii) what they observed 
are merely 
{\it equilibrium} position fluctuations of the particle.
Also, the Nos\'e-Hoover dynamics is not 
necessary to explain their results.

When a colloidal particle in water is trapped by a harmonic optical 
trap moving at a 
constant velocity $v_0$, the position along the moving direction, $x$, 
obeys the Langevin equation 
\begin{equation}
\zeta {dx\over dt}+k(x-v_0 t) = f(t)
\label{EqM}
\end{equation}
where $\zeta$ is the drag coefficient and $k$ the trapping constant. 
The Langevin force $f(t)$ is a 
zero-mean Gaussian random variable satisfying $
\langle f(t) f(t')\rangle = 2 k_B T
\zeta \delta(t-t')$, 
where $T$ is the water temperature.
Then the position measured from the trap center, $\delta x(s)\equiv 
x(s) -v_0 s$, is
\begin{eqnarray}
\delta x(t) &=&  {\zeta v_0 (e^{-{k t\over
\zeta}}-1)\over
k}+x_0
e^{-{kt\over
\zeta}} +\int_0^t {ds \over \zeta}
e^{-{k(t-s)\over
\zeta}} f(s)
\label{Solution}
\end{eqnarray}
where the initial position $x_0$ ia a random variable obeying an 
{\it equilibrium}
distribution
$\rho_{\rm eq}(x_0)\propto 
\exp\left({-k \ x_0^2\over 2k_BT}\right)$.

In Ref. 1, the work done by the
moving trap divided by 
$k_BT$, $\Sigma_t \equiv {1\over k_B T}
\int_0^t ds \ v_0 \cdot \{ -k \delta x(s) \}$, is investigated for 
individual trajectories.
As a linear combination of mutually independent Gaussian 
random variables, $\Sigma_t$ is Gaussian
with average
$m_t
\equiv
\langle
\Sigma_t
\rangle$ and variance
$\sigma_t^2
\equiv \langle (\Sigma_t-m_t)^2
\rangle$.
From (\ref{Solution}), one obtains
$\sigma_t^2 = 2 m_t$,
\begin{eqnarray}
m_t = {v_0^2 \zeta \over k_B T} \left\{
t- {\zeta \over k}\left(1-e^{-{kt\over
\zeta}}\right)\right\} \label{average}
\end{eqnarray}
then, `the 
transient fluctuation theorem' of Ref. 1
\begin{eqnarray}
{{\rm Pr}(\Sigma_t <0)\over {\rm Pr}(\Sigma_t>0)}
&=& \left\langle \exp\left(-{2m_t \over
\sigma_t^2}\Sigma_t
\right)\right\rangle_{\Sigma_t>0}
=\left\langle
e^{-\Sigma_t}\right\rangle_{\Sigma_t>0}
\nonumber
\end{eqnarray}
and an explicit expression in terms of the error function erf($x$)
\cite{Abramo}
\begin{equation}
\left\langle
e^{-\Sigma_t}\right\rangle_{\Sigma_t>0}
= {1-{\rm erf}(\sqrt{m_t}/2)\over
1+{\rm erf}(\sqrt{m_t}/2)}
\label{Result}
\end{equation}
In Fig. 1, Eq.(\ref{Result}) and the result
of Ref. 1 are compared for parameter values
consistent with the experiment. 
The agreement is quite satisfactory.

Although the colloidal particle sometimes does work because of 
${\rm Pr}(\Sigma_t <0)\not= 0$, it does not do the net work 
as an average: $\langle \Sigma_t \rangle =m_t \ge 0$ and
the second law is  not violated.
In this respect, the situation is similar to that in the Feynman 
ratchet\cite{Feynman} at thermal equilibrium, where the ratchet 
never does a net work in spite of fluctuations and whose behavior
is consistent with the second law.

As easily seen, 
the dominant contribution comes from the $x_0$-fluctuation. 
This means that the `entropy consuming trajectories', 
for which $\Sigma_t <0$, are generated by {\it equilibrium} 
fluctuations: 
Suppose, just before the trap starts to move, 
thermal fluctuations cause a particle translation from the trap center 
along 
the direction of the trap motion. Then, even after the trap starts to 
move, 
the particle moves backwards because of the harmonic force from the trap
and $\Sigma_t$ is negative. 
This process continues until the particle reaches the new equilibrium 
position.
In short, the trajectories with negative $\Sigma_t$ are merely the 
trajectories
whose 
initial positions 
happen to translate opposite to the trap motion by {\it equilibrium 
fluctuation}. 
\begin{figure}[h]
\epsfxsize=6.5cm
\centerline{\epsfbox{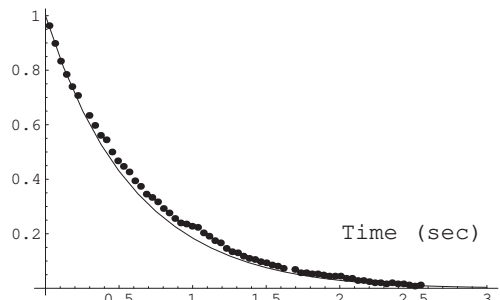}}
\caption{ 
Eq.(\ref{Result}) (solid curve) v.s. the
experimental result of Ref. 1 (dots). 
The Stokes law for the slip boundary condition: $\zeta =4\pi a \eta$
is assumed with the particle radius $a=3.15 \mu$m and the water 
viscosity 
$\eta$ at $T=303$ K. The
value $v_0=1.25  \mu$m/s is given
in Ref. 1 and $\zeta/k=$ 2.5 sec is assumed.
}
\end{figure}


\vskip -20pt
\begin{thebibliography}{99}
\bibitem{WSMSE} G.M. Wang, E.M. Sevick,
E. Mittag, D.J. Searles and D.J. Evans, 
Phys. Rev. Lett., {\bf 89}, 050601 (2002)

\bibitem{Abramo} M. Abramowitz and I.A. Stegun,
{\it Handbook of Mathematical Functions}, (Dover,
New York, 1972).

\bibitem{Feynman} R.P. Feynman, R.B. Leighton and M. Sands,
{\it The Feynman Lectures on Physics} vol.1, (Addison-Wesley,
Reading, 1963). See also T.M. Nieuwenhuizen and A.E.
Allahverdyan, cond-mat/0207587 and references
therein.


\end{thebibliography}
\end{document}